 \documentclass[final,5p,times,twocolumn]{elsarticle} 

\usepackage{amssymb}
\usepackage{lipsum}
\usepackage{amsmath}
\usepackage{bm}
\usepackage{hyperref}

\usepackage{color}

\biboptions{sort&compress}

\journal{Physics Letters B} 

\begin{document}
\begin{frontmatter}

\title{Wigner and mirror correlations in nuclear mass predictions with kernel ridge regression}

\author[FZU]{X. H. Wu}
\ead{wuxinhui@fzu.edu.cn}
\address[FZU]{Department of Physics, Fuzhou University, Fuzhou 350108, Fujian, China}

\begin{abstract}
Wigner- and mirror-correlated kernel ridge regression (WKRR) and its odd-even extension WKRRoe are developed to improve nuclear mass predictions.
The Wigner and mirror correlations are encoded entirely in the kernel functions without introducing more weight parameters.
For experimentally known nuclear masses, WKRRoe achieves a leave-one-out root-mean-square deviation of 98.0 keV, which falls below the often-quoted 100-keV scale discussed in connection with chaos-related limits.
The gain is mainly concentrated in light nuclei near $N=Z$ and is especially large for mirror pairs.
Further tests show that WKRRoe can potentially improve predictions for experimentally unknown nuclei near $N=Z$, particularly when their mirror partners have already been measured.
\end{abstract}


\end{frontmatter}

\section{Introduction}

Nuclear masses serve as basic nuclear quantities that integrate shell structure, pairing, deformation, and isospin-dependent correlations.
Accurate masses are needed to delimit nuclear stability, quantify separation and decay energies, and provide nuclear input for nucleosynthesis calculations~\cite{Lunney2003RMP, Mumpower2016PPNP, Jiang2021Astrophys.J.}.
Advances in radioactive-ion-beam facilities have enabled about 2500 nuclear masses to be evaluated experimentally, but large neutron-rich and proton-rich regions, especially toward the drip lines, remain beyond present experimental reach~\cite{Wang2021AME2020, Yamaguchi2021PPNP}.

Global nuclear mass predictions must cover thousands of experimentally inaccessible nuclei, so a useful approach must combine accurate interpolation of known masses with controlled behavior away from data.
Modern global models can reproduce the gross mass surface but retain local and regional residual structures in different ways.
Representative descriptions include macroscopic-microscopic models, such as the finite-range droplet model~\cite{Moeller2016FRDM} and the Weizs\"acker-Skyrme model~\cite{Wang2014WS4}, phenomenological and shell-model-inspired formulas such as KTUY~\cite{Koura2005KTUY} and Duflo-Zuker~\cite{Duflo1995DZ}, and microscopic mass tables and their underlying frameworks based on nonrelativistic energy density functionals~\cite{Goriely2009HFB17,Goriely2009D1M,Goriely2013HFB,Erler2012Nature} and covariant density functionals~\cite{Geng2005RMF,PenaArteaga2016RMF,Afanasjev2013CDFT,Xia2018RCHB,Yang2021CDFT,Zhang2022DRHBc, Pan2022DRHBc,Guo2024ADNDT}.
These frameworks incorporate bulk, shell, pairing, deformation, and continuum effects at different levels and with different approximations, and the physics that remains missing or imperfectly described can therefore be model dependent.
Indeed, recent principal-component analyses of nuclear mass predictions~\cite{Wu2025SCPMA} and residuals~\cite{Huang2026PCA} found no single dominant residual pattern common to all models; instead, the residual structures are largely model specific, suggesting that different models miss distinct combinations of nuclear effects.

Data-driven residual corrections can learn such missing or imperfectly described effects and thus improve nuclear mass predictions.
A variety of machine-learning approaches have been explored for this purpose, including radial-basis-function methods~\cite{Wang2011RBF, Niu2013RBF, Li2026RBFms}, Bayesian, feed-forward, and convolutional neural networks~\cite{Utama2016PRC, Niu2022BML, Lu2025CNN}, physically informed machine-learning models~\cite{Mumpower2022PIML, Li2024MLrprocess, Bentley2025PIML}, Gaussian-process regression and Bayesian model averaging~\cite{Neufcourt2019BMA, Shelley2021GPR, Huang2026GP}, and kernel ridge regression (KRR)~\cite{Wu2020KRR}.
Among them, KRR is a regularized nonlinear regression method in which physically motivated features can be incorporated directly into the kernel function, thereby making the learned correlations comparatively transparent and interpretable.

In KRR, the kernel quantifies the relevance of one nucleus to another.
Consequently, changing the kernel is not merely a numerical adjustment; it is a statement about which nuclear correlations should be transferred through the data.
The first application of KRR to nuclear masses used a Gaussian kernel in the two-dimensional coordinate $\bm{x}=(Z,N)$~\cite{Wu2020KRR}.
Nearby nuclei then have large similarity, whereas corrections vanish continuously when the distance from all training nuclei grows.
The KRRoe extension augmented the kernel so that nuclei with the same proton and neutron parities share an additional, longer-ranged correlation~\cite{Wu2021KRRoe}.
This separates the four odd-even sublattices and greatly improves both masses and one-nucleon separation energies.
Subsequent developments further demonstrate how physical relations can be implemented at the kernel level.
Gradient KRR couples masses and their finite differences in a multitask kernel~\cite{Wu2022GKRR}; anisotropic KRR strengthens correlations along isotope and isotone directions~\cite{Wu2024AKRR}; and systematic kernel comparisons have clarified the role of locality and asymptotic decay~\cite{Wu2023Kernels}.
More recently, complementary mass-derived observables have been incorporated into anisotropic KRR to improve extrapolative predictions~\cite{Tian2025AKRRCMO}.
The same framework has also been used to refine relativistic continuum Hartree-Bogoliubov mass tables, examine the physical structures learned by KRR, and assess the impact of corrected masses on $r$-process simulations~\cite{Guo2022KRRrprocess, Du2023KRRPhysics, Wu2024RCHBKRR, Guo2024RCHBKRRoe}.
The successful applications of KRR to nuclear masses have also stimulated its use in other areas of nuclear physics, including nuclear energy density functionals~\cite{Wu2022Phys.Rev.C, Chen2024IJMPE, Wu2025CP}, charge radii~\cite{Ma2022Chin.Phys.C, Tang2024NST}, and neutron-capture cross sections~\cite{Huang2022Commun.Theor.Phys.}.
These studies motivate a general principle in the KRR framework, i.e., improvements should follow from a physically identifiable modification of nuclear similarity and should be tested against the correlations that motivated it.

Two such correlations of nuclear similarity can be further considered.
First, experimental mass surfaces show a cusp associated with the Wigner energy, conventionally connected with enhanced neutron-proton correlations at small isospin~\cite{Satula1997Wigner, Negrea2014Wigner}.
The relative asymmetry $I=\frac{|N-Z|}{A}$,
therefore provides a natural measure of proximity to the Wigner region.
Second, approximate charge symmetry relates a nucleus $(Z,N)$ to its mirror $(N,Z)$ after Coulomb and other isospin-breaking contributions are accounted for~\cite{Miller1990CSB}.
Mirror constraints and empirical mirror-mass relations have long been used to improve mass estimates~\cite{Wang2010Mirror, Bao2016Mirror, Zong2019Mirror, Ma2020MirrorLocal, Guo2024Mirror, Wang2025Mirror}.
Taken together, these observations imply that nuclear similarity should be enhanced both among nearby nuclei with comparable Wigner coordinates and between a nucleus and the neighborhood of its mirror image, thereby providing a direct physical guide for redesigning the KRR kernel.

In this work, we incorporate the Wigner and mirror correlations into the KRR framework to improve the description of nuclear masses based on the WS4 model.
Both correlations are introduced through the ingenious designs of kernel function by using a multiplicative similarity factor in $I$ to select the Wigner region and a reflected channel to connect $(Z,N)$ to the neighborhood of $(N,Z)$.
The construction requires no additional input variables or weight parameters outside the kernel in the KRR framework.

\section{Theoretical framework}
\label{sec:framework}

Following Refs.~\cite{Wu2020KRR,Wu2021KRRoe}, KRR and KRR with odd-even effects (KRRoe) learn the mass residual $y(\bm{x}_i)=M_{\mathrm{exp}}(\bm{x}_i)-M_{\mathrm{th}}(\bm{x}_i)$ at $\bm{x}_i=(Z_i,N_i)$.
Both can be written as
\begin{align}
S(\bm{x}_j)={}&\sum_{i=1}^{m}K'(\bm{x}_j,\bm{x}_i)\alpha_i,
\label{eq:krr_function_effective}\\
\bm{\alpha}={}&(\bm{K}'+\lambda\bm{I})^{-1}\bm{y},
\label{eq:alpha}
\end{align}
where $m$ is the number of training nuclei, $\bm{K}'$ is the kernel matrix, $\bm{I}$ is the identity matrix, $\lambda$ is the regularization strength, and $\alpha_i$'s are weight parameters.
For KRR, the kernel $K'$ reads
\begin{equation}
K'(\bm{x}_j,\bm{x}_i)=
\exp\left[-\frac{|\bm{x}_i-\bm{x}_j|^2}{2\sigma^2}\right],
\label{eq:krr_kernel}
\end{equation}
where $|\cdot|$ denotes the Euclidean norm.
KRRoe accounts for the four pairing-induced odd-even classes by adding
\begin{equation}
K_{\mathrm{oe}}(\bm{x}_j,\bm{x}_i)=
\delta_{\mathrm{oe}}
\exp\left[-\frac{|\bm{x}_i-\bm{x}_j|^2}{2\sigma_{\mathrm{oe}}^2}\right].
\label{eq:krr_kernel_oe}
\end{equation}
Here and below, $\delta_{\mathrm{oe}}$ is the odd-even selector for the coordinate pair, which is 1 when both proton and neutron parities agree and zero otherwise. $\sigma_{\mathrm{oe}}$ is the odd-even length scale.
The full kernel used in the KRRoe approach is
\begin{equation}
K'(\bm{x}_j,\bm{x}_i)=K(\bm{x}_j,\bm{x}_i)
+\frac{\lambda}{\lambda_{\mathrm{oe}}}
K_{\mathrm{oe}}(\bm{x}_j,\bm{x}_i).
\label{eq:krr_kernel_effective}
\end{equation}
Here, $\lambda_{\mathrm{oe}}$ is the regularization parameter for the odd-even part.

\begin{figure}[!t]
\centering
\includegraphics[width=\columnwidth]{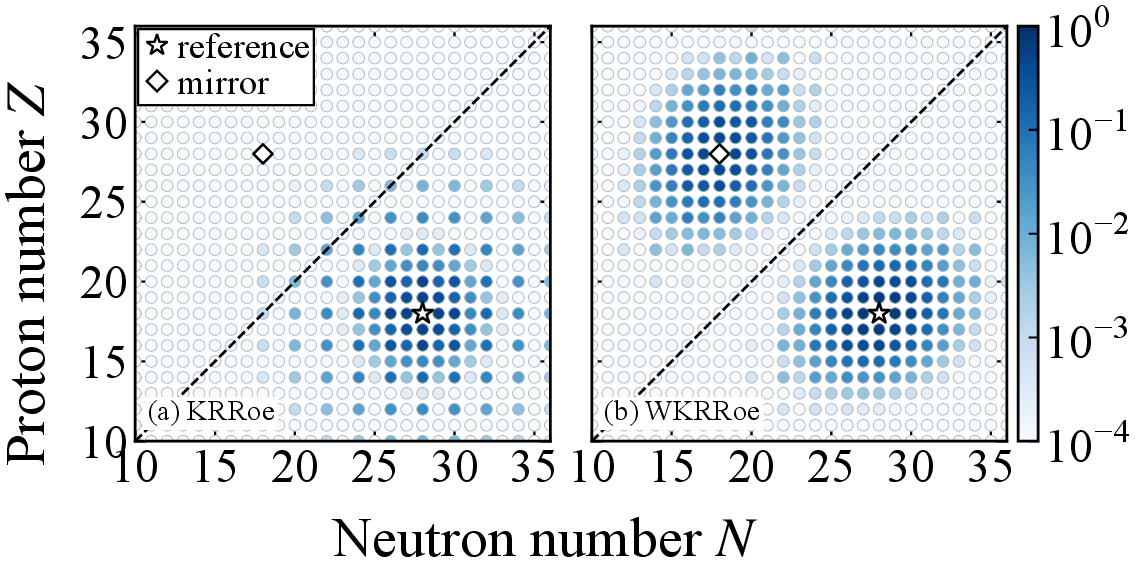}
\caption{Normalized kernels relative to reference nucleus $\bm{x}_0=(18,28)$ for (a) KRRoe and (b) WKRRoe. Color gives $K'(\bm{x},\bm{x}_0)/K'(\bm{x}_0,\bm{x}_0)$; the star, open diamond, and dashed line mark $\bm{x}_0$, its mirror, and $N=Z$, respectively. The kernel hyperparameters are the optimized values as shown in Sec.~\ref{sec:numerics} for numerical details.}
\label{fig:kernel}
\end{figure}

The Wigner and mirror correlations are brought into the KRR approach in the following way.
Using relative asymmetry $I$, Wigner- and mirror-correlated KRR (WKRR) replaces Eq.~\eqref{eq:krr_kernel} by
\begin{multline}
K'(\bm{x}_j,\bm{x}_i)=
\Bigg\{\exp\left[-\frac{|\bm{x}_i-\bm{x}_j|^2}{2\sigma^2}\right]
\\
+\rho\exp\left[-\frac{|\widetilde{\bm{x}}_i-\bm{x}_j|^2}{2\sigma^2}\right]\Bigg\}
\exp\left[-\frac{(I_i-I_j)^2}{2\sigma_I^2}\right].
\label{eq:wkrr}
\end{multline}
Here, $\widetilde{\bm{x}}_i=(N_i,Z_i)$ is the mirror coordinate of nucleus $i$, defined by interchanging its proton and neutron numbers.
The last factor favors similar $I$, while the term centered at $\widetilde{\bm{x}}_i$ transfers mirror information; $\sigma_I$ and $\rho$ set the Wigner range and mirror strength, respectively.
WKRR has no parity-selective channel.
Its odd-even extension WKRRoe uses Eq.~\eqref{eq:wkrr} together with
\begin{multline}
K_{\mathrm{oe}}(\bm{x}_j,\bm{x}_i)=
\delta_{\mathrm{oe}}\!
\exp\left[-\frac{|\bm{x}_i-\bm{x}_j|^2}{2\sigma_{\mathrm{oe}}^2}\right]
\!
\exp\left[-\frac{(I_i-I_j)^2}{2\sigma_{I,\mathrm{oe}}^2}\right]
\\
+\rho_{\mathrm{oe}}\delta_{\mathrm{oe}}\!
\exp\left[-\frac{|\widetilde{\bm{x}}_i-\bm{x}_j|^2}{2\sigma_{\mathrm{oe}}^2}\right]
\!
\exp\left[-\frac{(I_i-I_j)^2}{2\sigma_{I,\mathrm{oe}}^2}\right].
\label{eq:wkrroe_oe}
\end{multline}
In the reflected term, $\delta_{\mathrm{oe}}$ applies to $\bm{x}_j$ and $\widetilde{\bm{x}}_i$, while $\rho_{\mathrm{oe}}$ and $\sigma_{I,\mathrm{oe}}$ set its mirror strength and $I$ range.
The prediction and coefficients retain Eqs.~\eqref{eq:krr_function_effective} and \eqref{eq:alpha}, and hence WKRRoe adds no independent weights.
The remodulated WKRR and WKRRoe kernels enhance correlations both between nuclei with similar $I$, corresponding to similar relative neutron-proton asymmetry, and between mirror nuclei.
Figure~\ref{fig:kernel} visualizes these two enhancements by comparing the kernel patterns of the KRRoe and WKRRoe approaches.
KRRoe shows only a local neighborhood around the reference nucleus [Fig.~\ref{fig:kernel}(a)], whereas WKRRoe favors nuclei with $I$ close to that of the reference and produces an additional localized neighborhood around its mirror [Fig.~\ref{fig:kernel}(b)]; the alternating pattern is generated by $\delta_{\mathrm{oe}}$.

\section{Numerical details}
\label{sec:numerics}

The baseline masses are taken from the WS4 predictions~\cite{Wang2014WS4}, and experimental mass excesses and uncertainties are taken from AME2020~\cite{Wang2021AME2020}.
The overlap satisfying $Z\geq8$, $N\geq8$, and an experimental uncertainty below 100 keV contains 2340 nuclei.

The hyperparameters of KRR, WKRR, KRRoe, and WKRRoe approaches are determined by minimizing the rms deviations obtained from leave-one-out (LOO) cross-validation~\cite{Wu2021KRRoe}, in which the mass of each nucleus is predicted using the corresponding model trained on all the other 2339 nuclei.
For KRR, the optimized hyperparameters are $\sigma=2.1773$ and $\lambda=0.2977$.
For WKRR, the optimized hyperparameters are $\sigma=2.1292$, $\lambda=0.4969$, $\sigma_I=0.0823$, and $\rho=0.8387$.
For KRRoe, the optimized hyperparameters are $\sigma=1.2845$, $\lambda=0.0659$, $\sigma_{\mathrm{oe}}=2.9494$, and $\lambda_{\mathrm{oe}}=0.1443$.
For WKRRoe, the optimized hyperparameters are $\sigma=1.8087$, $\lambda=0.0143$, $\sigma_{\mathrm{oe}}=2.6872$, $\lambda_{\mathrm{oe}}=0.0662$, $\sigma_I=0.0782$, $\sigma_{I,\mathrm{oe}}=0.0536$, $\rho=0.6199$, and $\rho_{\mathrm{oe}}=0.9441$.

\section{Results and discussion}
\label{sec:results}

\begin{figure}[htbp]
 \centering
 \includegraphics[width=0.9\columnwidth]{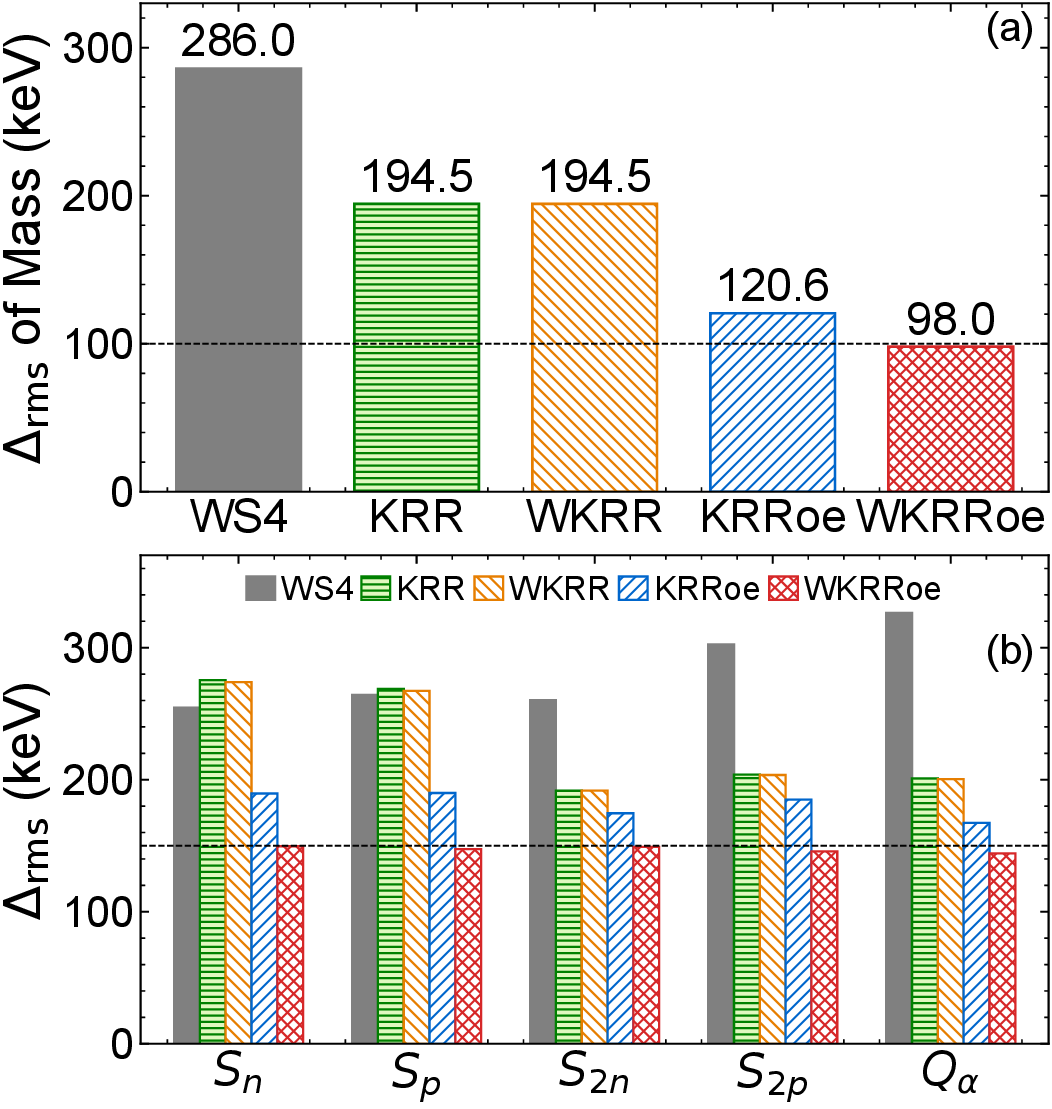}
 \caption{Root-mean-square deviations obtained from LOO predictions for the 2340-nucleus data set. Panel (a) compares the masses from WS4 and the four kernel corrections. Panel (b) shows one- and two-nucleon separation energies and $\alpha$-decay energies derived from the same LOO masses. The dashed lines mark 100 and 150 keV, respectively.}
 \label{fig:loorms}
\end{figure}

Figure~\ref{fig:loorms}(a) compares the mass prediction accuracies of different models.
The rms deviation of WS4 is 286.0 keV and is reduced to 194.5 keV by KRR.
Introducing the odd-even kernel gives a further substantial decrease to 120.6 keV for KRRoe.
In contrast, WKRR yields 194.5 keV, essentially the same value as KRR, whereas WKRRoe reaches 98.0 keV.
Thus, the Wigner and mirror correlations improve the mass description only after the odd-even sublattices are resolved.
The resulting accuracy of 98.0 keV also crosses the often-quoted 100-keV benchmark, which has been discussed as an approximate upper bound on possible chaos-related unpredictability in nuclear mass predictions~\cite{Bohigas2002Chaos, Barea2005Chaos, Niu2018SciBull, Niu2022BML}.

Figure~\ref{fig:loorms}(b) examines whether the improved masses also define a more consistent local mass surface.
The separation energies and $Q_\alpha$ values are calculated from the nuclear masses and are not included as training targets.
It can be seen that WKRRoe gives the smallest rms deviation for all five quantities.
Its rms deviations are 149.6, 147.5, 149.1, 145.7, and 144.3 keV for $S_n$, $S_p$, $S_{2n}$, $S_{2p}$, and $Q_\alpha$, respectively, all below 150 keV.
The gain is therefore not limited to the nuclear masses but survives the finite differences relevant to separation and decay energies.

\begin{figure}[htbp]
 \centering
 \includegraphics[width=0.9\columnwidth]{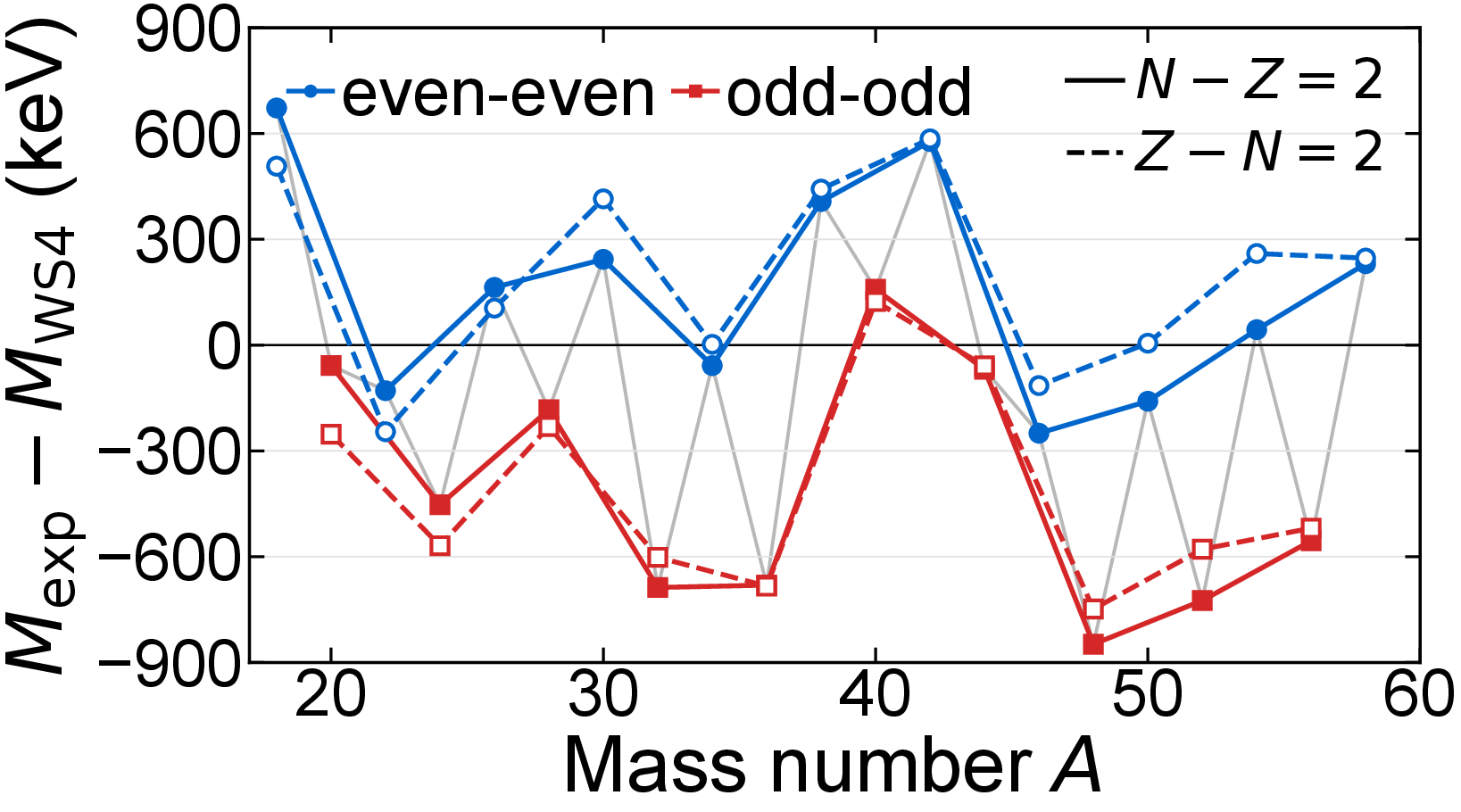}
 \caption{WS4 mass residuals for 21 measured mirror pairs on the $N-Z=2$ and $Z-N=2$ lines. Blue circles and red squares denote even-even and odd-odd nuclei, respectively. Solid filled and dashed open symbols distinguish the neutron-rich and proton-rich members. The thin gray line joins consecutive nuclei along $N-Z=2$ without separating their odd-even classes.}
 \label{fig:wmexplain}
\end{figure}

Figure~\ref{fig:wmexplain} helps explain why WKRR provides no improvement over KRR, whereas WKRRoe improves upon KRRoe.
The $|N-Z|=2$ chains are particularly instructive because they contain measured mirror pairs close to the Wigner region and alternate between even-even and odd-odd nuclei.
If the $N-Z=2$ nuclei are followed without resolving parity, the gray trajectory changes abruptly from one nucleus to the next.
The blue and red points instead form distinct parity branches, and the solid and dashed members of many mirror pairs follow the same branch.
Thus, the raw residual field contains two simultaneous structures: pairing separates neighboring nuclei, whereas Wigner and mirror symmetry relates nuclei within the appropriate odd-even class.
A Wigner-modulated kernel without the odd-even gate mixes these branches, and thus learns almost no net benefit.
KRRoe first confines this transfer to an odd-even sublattice, where the residual variation is smoother and the Wigner and mirror relations become learnable.
Therefore, odd-even resolution is necessary, although it does not by itself remove the remaining Wigner and mirror structure.

\begin{figure}[htbp]
 \centering
 \includegraphics[width=0.9\columnwidth]{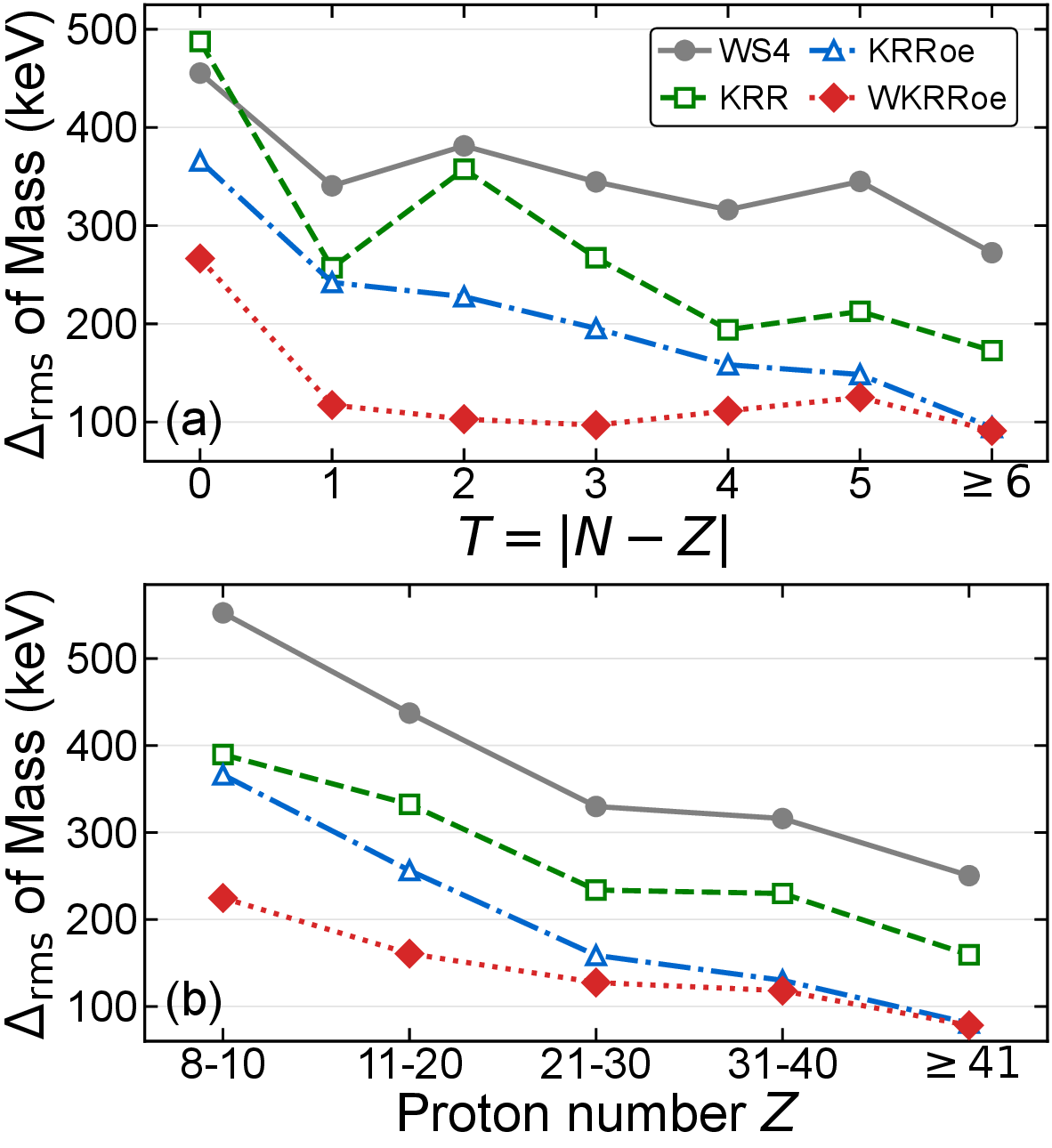}
 \caption{Mass rms deviations grouped by (a) $T=|N-Z|$ and (b) proton number $Z$. The nuclei are divided into bins according to their locations on the nuclear chart, and each point is evaluated from all nuclei in the corresponding bin.}
 \label{fig:wigner}
\end{figure}

Figure~\ref{fig:wigner} presents the WKRRoe improvement for nuclei in different regions of the nuclear chart.
Panel (a) groups the LOO deviations by $T=|N-Z|$, which denotes the integer distance from the $N=Z$ line.
For $T=0$, 1, 2, and 3, WKRRoe significantly reduces the KRRoe deviation by 99.3, 124.9, 124.7, and 98.5 keV, respectively.
The reduction decreases to 46.9 keV at $T=4$, 23.2 keV at $T=5$, and only 3.5 keV for $T\geq6$.
The effect is consequently concentrated near $N=Z$, as expected for a kernel designed to represent Wigner-related correlations, rather than being a uniform readjustment of KRRoe.

Panel (b) shows the associated dependence on proton number.
The KRRoe-to-WKRRoe reduction is 141.5 keV for $8\leq Z\leq10$ and 95.5 keV for $11\leq Z\leq20$, decreases to 31.3 keV for $21\leq Z\leq30$, and 12.0 keV for $31\leq Z\leq40$, and is only 2.6 keV for $Z\geq41$.
This strong light-nucleus dependence follows from the experimental distribution, i.e., nuclei close to proton-neutron symmetry occur predominantly at low $Z$, while medium-heavy measured nuclei lie farther from $N=Z$.
This localization is also physically consistent with established Wigner-energy systematics.
Empirical mass analyses associate the Wigner cusp with the additional binding near $N=Z$~\cite{Satula1997Wigner}, find an approximately inverse-mass dependence of its coefficient~\cite{Cheng2015Wigner}, and show that additional Wigner terms confined to light nuclei can significantly improve global mass fits~\cite{Lunney2003RMP}.

\begin{figure}[htbp]
 \centering
 \includegraphics[width=0.9\columnwidth]{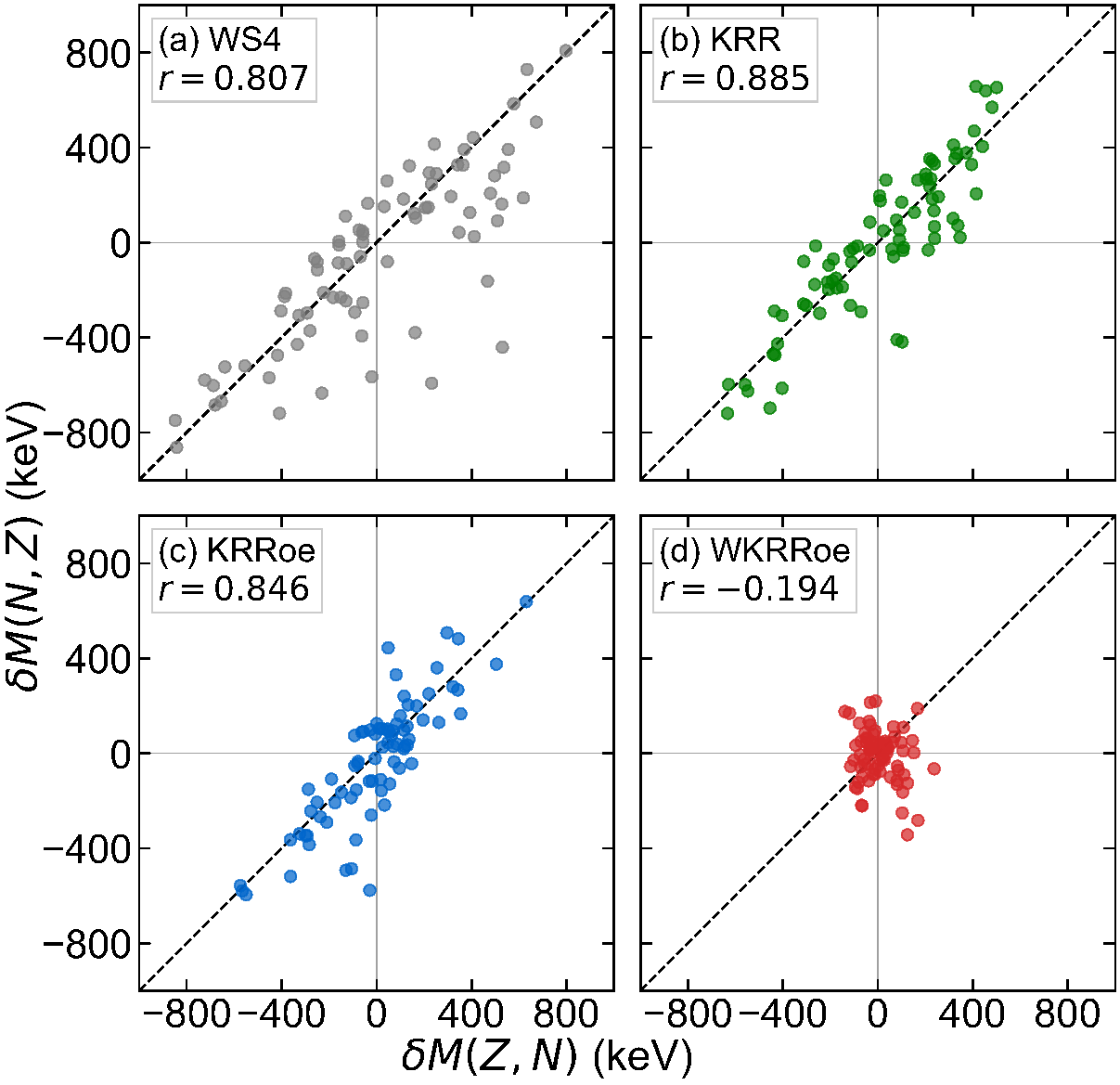}
 \caption{Correlations between the mass residuals of 77 measured mirror pairs. For each pair, $(Z,N)$ denotes the member with $N>Z$ and $(N,Z)$ its mirror partner; the abscissa and ordinate are their respective residuals $\delta M=M_{\mathrm{exp}}-M_{\mathrm{pred}}$. Panels (a)-(d) show WS4, KRR, KRRoe, and WKRRoe, respectively. The dashed line denotes equal residuals for the two partners, and $r$ is the Pearson correlation coefficient.}
 \label{fig:mirror}
\end{figure}

The reflected kernel is motivated by a more specific relation, i.e., the residuals of mirror nuclei should contain a common component that a kernel centered only on the target coordinate cannot transfer directly~\cite{Ma2020MirrorLocal}.
Figure~\ref{fig:mirror} tests this premise for the 77 measured mirror pairs.
Across the first three panels, the residual cloud contracts but retains a pronounced alignment along the diagonal.
For the 154 nuclei belonging to these pairs, the rms deviation decreases from 383.3 keV for WS4 to 308.5 keV for KRR and 245.8 keV for KRRoe, while the corresponding correlations remain strong, with $r=0.807$, 0.885, and 0.846, respectively.
The larger value of $r$ after KRR does not imply lower accuracy, because the Pearson coefficient measures co-variation rather than residual magnitude.
Instead, the persistent correlation shows that KRR and KRRoe reduce other components of the error but leave a systematic component shared by mirror partners.

With WKRRoe, the residual cloud contracts further toward the origin, the rms deviation falls to 94.3 keV, and the correlation is reduced to $r=-0.194$.
The simultaneous reduction of the residual magnitude and the diagonal correlation shows that the reflected channel removes this common mirror error rather than merely rescaling it.
This result directly demonstrates the availability of mirror-to-target information transfer.

\begin{figure}[htbp]
 \centering
 \includegraphics[width=0.9\columnwidth]{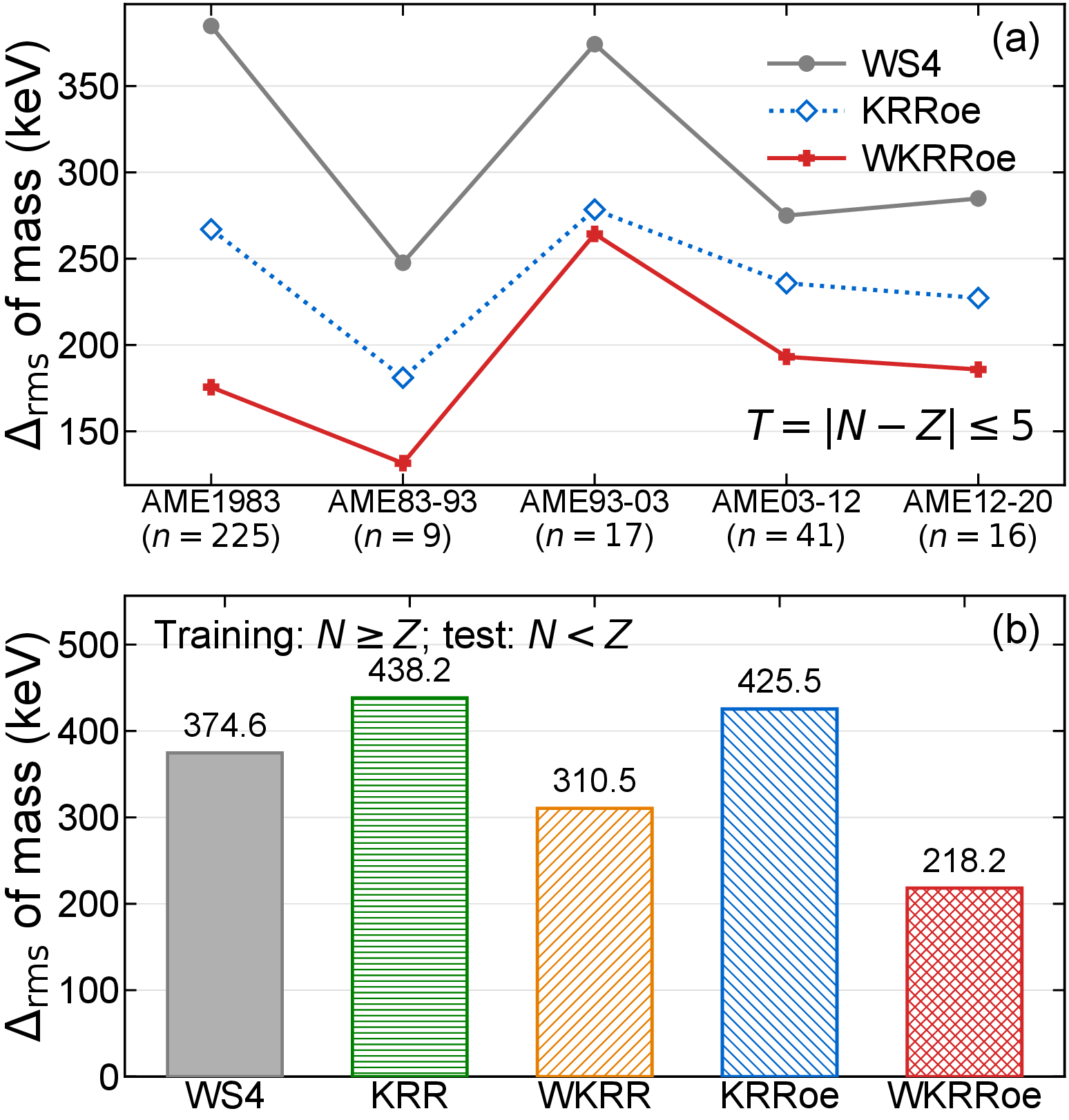}
 \caption{Extrapolation tests. Panel (a) shows rms deviations for $T\leq5$ nuclei. The AME1983 points are LOO predictions obtained within AME1983, whereas each subsequent group contains nuclei newly added between the indicated mass evaluations and is predicted by models trained only on AME1983. The sample size is printed below each group. Panel (b) shows an asymmetric mirror-transfer test in which all nuclei with $N<Z$ are withheld as the test set and the models are trained only on nuclei with $N\geq Z$.}
 \label{fig:extrapolation}
\end{figure}

The LOO results quantify interpolation among known masses but do not alone establish extrapolative reliability.
Following the mass-evaluation-sequence test used in Ref.~\cite{Guo2024RCHBKRRoe}, Fig.~\ref{fig:extrapolation}(a) tests extrapolation in the Wigner-sensitive region $T\leq5$ by training the models on the masses included in AME1983~\cite{Wapstra1985AME1983} and predicting nuclei first included in AME1993, AME2003, AME2012, and AME2020~\cite{Audi1993AME1993,Audi2003AME2003,Wang2012AME2012,Wang2021AME2020}.
For the AME1983 set, the plotted values are LOO results obtained within that data set; each group of nuclei added in a subsequent mass evaluation is predicted by models trained only on AME1983.
WKRRoe is more accurate than KRRoe for every subsequently added group: the deviations change from 181.0 to 131.5 keV for AME83-93, from 278.4 to 264.3 keV for AME93-03, from 235.8 to 193.1 keV for AME03-12, and from 227.2 to 185.8 keV for AME12-20.
These extrapolation results show that WKRRoe improves mass predictions beyond its training data and can therefore be useful for predicting experimentally unknown nuclei near the $N=Z$ line.

Panel (b) provides a complementary asymmetric test of mirror transfer.
All nuclei with $N<Z$ are withheld as the test set, while only nuclei with $N\geq Z$ are used for training.
In this test, ordinary KRR and KRRoe give predictions that are even less accurate than those of WS4, with rms deviations of 438.2 and 425.5 keV, respectively, compared with 374.6 keV for WS4.
The mirror-aware models behave differently, i.e., WKRR gives 310.5 keV and parity-resolved WKRRoe reaches 218.2 keV.
This shows that WKRRoe is particularly useful for predicting masses of experimentally unknown nuclei when the related mirror partners are known.

\section{Summary}
\label{sec:summary}

The Wigner- and mirror-correlated KRR (WKRR) and its odd-even extension WKRRoe for correcting the WS4 nuclear masses have been developed.
The Wigner correlation is introduced through a multiplicative Gaussian in $I=|N-Z|/A$, while the mirror correlation is incorporated through a reflected Gaussian centered at the mirror coordinate.
Both correlations are implemented entirely through the kernel design, without introducing additional input variables or weight parameters outside the KRR kernel.

For the 2340 nuclei shared by WS4 and AME2020, the LOO rms mass deviations are 286.0, 194.5, 194.5, 120.6, and 98.0 keV for WS4, KRR, WKRR, KRRoe, and WKRRoe, respectively.
Note that the WKRRoe achieves an rms mass accuracy better than 100 keV.
The comparison between WKRR and WKRRoe shows that the Wigner and mirror correlations improve the mass description only after the odd-even sublattices are resolved.
WKRRoe also gives the smallest rms deviations for $S_n$, $S_p$, $S_{2n}$, $S_{2p}$, and $Q_\alpha$, all below 150 keV.
The improvement is concentrated in light nuclei and near the $N=Z$ line, consistent with established Wigner-energy systematics.
For the 77 measured mirror pairs, WKRRoe reduces the residual correlation from $r=0.846$ to $r=-0.194$ and reduces the rms deviation to 94.3 keV, demonstrating effective information transfer from a measured mirror partner.

The extrapolation tests further show that these improvements are not restricted to LOO interpolation.
When trained only on the AME1983 masses, WKRRoe outperforms KRRoe for every group of $T\leq5$ nuclei added in subsequent mass evaluations.
In the asymmetric test trained on nuclei with $N\geq Z$ and evaluated on those with $N<Z$, WKRRoe reaches an rms deviation of 218.2 keV, compared with 425.5 keV for KRRoe and 374.6 keV for WS4.
These results demonstrate that incorporating physically motivated Wigner and mirror correlations into the kernel provides an accurate and interpretable approach to nuclear mass prediction, particularly for experimentally unknown nuclei near $N=Z$ and for nuclei whose mirror partners have already been measured.

\section*{Acknowledgments}
This work was supported by the National Natural Science Foundation of China under Grant No. 12405134.

\bibliographystyle{elsarticle-num}
\bibliography{paper}

\end{document}